\documentclass{emulateapj}
\slugcomment{Accepted to The Astrophysical Journal Supplement}
\usepackage{apjfonts}

\usepackage{natbib}
\usepackage{amsmath}
\usepackage{lscape}
\usepackage{longtable}
\usepackage{multirow}
\usepackage{hyperref}

\newcommand{\ra}{\mathrm{RA}}
\newcommand{\dec}{\mathrm{dec}}

\newcommand{\beq}{\begin{equation}}
\newcommand{\eeq}{\end{equation}}

\newcommand{\subscript}[1]{\ensuremath{_{\textrm{#1}}}}

\begin{document}

\title{Extended Photometry for the DEEP2 Galaxy Redshift Survey: A Testbed for Photometric Redshift Experiments}
\author{Daniel J. Matthews\altaffilmark{1}, Jeffrey A. Newman\altaffilmark{1}, Alison L. Coil\altaffilmark{2}, Michael C. Cooper\altaffilmark{3} and Stephen D.J. Gwyn\altaffilmark{4}}
\altaffiltext{1}{Department of Physics and Astronomy, University of Pittsburgh, 3941 O'Hara Street, Pittsburgh, PA 15260}
\altaffiltext{2}{Department of Physics, University of California San Diego, 9500 Gilman Dr., La Jolla, CA 92093}
\altaffiltext{3}{Center for Galaxy Evolution, Department of Physics and Astronomy, University of California, Irvine, 4129 Frederick Reines Hall, Irvine, CA 92697, USA}
\altaffiltext{4}{Canadian Astronomical Data Centre, Herzberg Institute of Astrophysics, 5071 West Saanich Road, Victoria, British Columbia, Canada V9E 2E7}
\email{djm70@pitt.edu, janewman@pitt.edu, acoil@ucsd.edu, m.cooper@uci.edu, Stephen.Gwyn@nrc-cnrc.gc.ca}

\begin{abstract}

This paper describes a new catalog that supplements the existing DEEP2 Galaxy Redshift Survey photometric and spectroscopic catalogs with $ugriz$ photometry from two other surveys; the Canada-France-Hawaii Legacy Survey (CFHTLS) and the Sloan Digital Sky Survey (SDSS). Each catalog is cross-matched by position on the sky in order to assign $ugriz$ photometry to objects in the DEEP2 catalogs. We have recalibrated the CFHTLS photometry where it overlaps DEEP2 in order to provide a more uniform dataset. We have also used this improved photometry to predict DEEP2 $BRI$ photometry in regions where only poorer measurements were available previously. In addition, we have included improved astrometry tied to SDSS rather than USNO-A2.0 for all DEEP2 objects. In total this catalog contains $\sim27,000$ objects with full $ugriz$ photometry as well as robust spectroscopic redshift measurements, $64\%$ of which have $r > 23$. By combining the secure and accurate redshifts of the DEEP2 Galaxy Redshift Survey with $ugriz$ photometry, we have created a catalog that can be used as an excellent testbed for future photo-$z$ studies, including tests of algorithms for surveys such as LSST and DES.

\end{abstract}

\keywords{Catalogs --- Surveys --- Galaxies: photometry --- Galaxies: distances and redshifts --- Techniques: photometric --- Astrometry}

\section{Introduction}
\label{sec:intro}
Future wide-area photometric surveys will obtain imaging for a very large number of galaxies ($\sim10^8-10^9$), and many of the cosmological measurements to be performed with this data will require redshift information for these objects. It is not feasible to measure spectroscopic redshifts for this many objects, mainly due to the integration time required to obtain spectra, and in addition, many of the galaxies are too faint for spectroscopy.  To meet this challenge, many techniques have been developed for estimating redshifts from photometric information, where the flux from the galaxy is measured in a few broadband filters. Because our knowledge of the true spectral energy distributions of the full range of galaxies is limited, a training set of objects with accurate spectroscopic redshifts is generally used to determine or refine relations between photometric observables and $z$ (e.g., \citet{1995AJ....110.2655C,2010ApJ...715..823G,2006A&A...457..841I}).  However, the combination of deep photometry in many bands with deep spectroscopy for calibration purposes is available in only a few fields.

The DEEP2 (Deep Extragalactic Evolutionary Probe 2) Galaxy Redshift Survey \citep{Newman} obtained secure and accurate redshifts for more than 38,000 objects in four widely separated fields. However, the photometry used for DEEP2 targeting was obtained in the $B$, $R$, and $I$ filters, while the deepest datasets to date utilize measurements in $ugriz$. We have now constructed a catalog combining DEEP2 spectroscopic redshifts with data from two $ugriz$ photometric surveys which have covered the same fields: the Canada-France-Hawaii Legacy Survey (CFHTLS) \citep{2012AJ....143...38G} and the Sloan Digital Sky Survey (SDSS) \citep{2012ApJS..203...21A,2009ApJS..182..543A}. In this paper, we present the details of this catalog and make it publicly available as a testbed for algorithm development for future photometric redshift studies. These catalogs can be downloaded at http://deep.ps.uci.edu/DR4/photo.extended.html.

In \S\ref{sec:datasets} we describe the three different data sets used in this paper. This new catalog also provides astrometry tied to SDSS (rather than USNO-A2.0) as a reference; the corrections required are described in \S\ref{sec:astrom}. In \S\ref{sec:deep2f1} we describe how we constructed the catalog for DEEP2 Field 1, commonly known as the Extended Groth Strip (EGS). In the course of this we  derive improved photometric calibrations for CFHTLS photometry in all pointings that overlap DEEP2 Field 1 (\S \ref{sec:cfhtcal}). We also determine color transformations between the DEEP2 $BRI$ and the CFHTLS $ugriz$ photometric systems for this field, allowing us to use CFHTLS data to predict $BRI$ magnitudes for a subset of DEEP2 objects which  had poorer measurements originally (\S\ref{sec:predphot}). In \S\ref{sec:deep2f234} we describe how we constructed the combined $ugriz$/redshift catalog for DEEP2 Fields 2, 3, and 4. In \S\ref{sec:datatables} we provide details of the parameters that are included in the resulting catalogs, and in \S\ref{sec:conclusion} we conclude and provide summary statistics for this new sample.

\section{Data Sets}
\label{sec:datasets}

The DEEP2 Galaxy Redshift Survey is a magnitude-limited spectroscopic survey of objects with $R_{AB}<24.1$ \citep{Newman}. Data was taken in four separate fields, with photometry in each field from CFHT 12K $BRI$ imaging (the {\it ``pcat''} catalogs). Subsets of each {\it pcat} catalog were targeted for spectroscopy in order to obtain redshifts (the {\it ``zcat''} catalogs). DEEP2 Field 1 is part of the Extended Groth Strip (EGS), where the {\it pcat} photometry was measured in four overlapping $0.5^{\circ} \times 0.7^{\circ}$ pointings of the 12K camera (labeled as pointings 11-14). For the DEEP2 spectroscopic survey in this field ({\it zcat} catalog), objects were targeted in a $0.25^\circ \times 2.0^\circ$ window which spans all four pointings. In DEEP2 Fields 2, 3, and 4, the {\it pcat} and {\it zcat} catalogs cover the same area on the sky, where data was taken in $0.5^\circ \times 2.0^\circ$ rectangular fields, with each field divided up into three separate pointings (labeled as 21-23, etc.). In Field 2, pointing 23 is not included in this catalog since it was not observed with the DEIMOS spectrograph in DEEP2 and also has inferior $BRI$ photometry. We include all of pointing 43 in Field 4 in this catalog although only part of this pointing was actually observed with DEIMOS and have redshifts. See \citet{2004ApJ...617..765C} and \citet{Newman} for details of both the {\it pcat} and {\it zcat} catalogs.

To provide $ugriz$ photometry for objects in DEEP2 Field 1, we used the publicly-available CFHTLS Wide Field $i$-band selected unified catalog, as well as the CFHTLS Deep Field $i$-band selected catalog \citep{2012AJ....143...38G} where it overlaps the DEEP2 pointings. Photometry was obtained using the wide field optical imaging camera MegaCam. We selected objects in the Wide catalog from the seven pointings that overlap DEEP2 Field 1 (each pointing $\sim0.9^{\circ} \times 0.9^{\circ}$), where each $ugriz$ band reaches depths of $u = 24.6-25.8,\  g = 26.0-26.4,\ r = 25.2-26.2,\ i = 24.7-25.2$, and $z = 23.8-24.8$ (span shows the range of depths over all seven pointings). Here we have defined the depth in each pointing as the magnitude at which the errors in each band correspond to a $5\sigma$ flux measurement. The CFHTLS Deep Field D3 ($\sim1.0^{\circ} \times 1.0^{\circ}$) partially overlaps DEEP2 pointings 11-13 and reaches depths of  $u = 27.1,\  g = 27.7,\ r = 27.5,\ i = 27.2$, and $z = 25.7$. For the $ugriz$ magnitudes we used the Kron-like elliptical aperture magnitudes designated by MAG\_AUTO in the catalog.

For $ugriz$ photometry in DEEP2 Fields 2-4 we used data from the SDSS catalogs. Where SDSS overlaps DEEP2 Field 2 we select both stars and galaxies that are flagged as having clean photometry in the DR9 data release \citep{2012ApJS..203...21A}. Where SDSS overlaps DEEP2 Fields 3 and 4 we select sources flagged as having clean photometry in Stripe 82, which goes deeper than typical SDSS fields due to co-adding repeated imaging scans (designated by runs 106 and 206 in the Stripe82 database in DR7) \citep{2009ApJS..182..543A}. In all three fields we use model magnitude photometry. The depths reached for DR9 (Stripe 82) objects that overlap DEEP2 Fields 1 and 2 (3 and 4) are given by $u = 21.6-22.1\ (23.3-23.5),\ g = 23.0-23.2\ (24.7-24.8),\ r = 22.7-23.1\ (24.3-24.5),\ i = 22.0-22.5\ (23.8-23.9),$ and $z = 20.5-20.9\ (22.0-22.4)$.

\section{Corrected astrometry}
\label{sec:astrom}

The DEEP2 astrometry measurements were determined using stars from the USNO-A2.0 system. The USNO-A2.0 astrometry contained a number of known systematic errors, which have been propagated into the DEEP2 astrometry.  Additionally, the {\tt imcat}-produced data reductions \citep{1999astro.ph..7229K,2011ascl.soft08001K} tend to have larger systematic astrometric errors at the edges of each pointing, presumably due to a lack of astrometric calibration stars beyond field edges.  The net result is that systematic astrometric errors vary over scales of $5-10\arcmin$, and can reach values of $\sim1\arcsec$ in the worst cases. It should be noted that these errors are referring to the absolute astrometry, and the relative astrometry at small scales ($\lesssim 1\arcmin$) will be much more accurate than this. For objects separated by more than $1\arcmin$, there will be systematic offsets in the relative astrometry increasing with separation. In addition, the public {\it pcat} catalogs for field 1 (EGS) that are available do include an astrometric correction that ties them to SDSS, and so the absolute astrometry is better than this in those catalogs. For consistency in these catalogs, we perform the same astrometric corrections in all fields, including field 1.

In order to allow improved comparisons to external catalogs, we have calculated corrected astrometry for each object in DEEP2 using the superior absolute astrometry from SDSS as a reference frame rather than USNO-A2.0. The SDSS astrometry is calibrated against the Second Data Release of USNO CCD Astrograph Catalog (UCAC2), which measured the positions and proper motions for millions of stars, where the precision of measured positions are $\sim15-70$ mas, with systematics estimated to be $< 10$ mas \citep{2004AJ....127.3043Z}. For consistency in the catalog, we also performed corrections on the CFHTLS astrometry. The size and direction of the deviations from the SDSS astrometry varied significantly across each DEEP2 pointing; therefore, it was necessary to calculate a correction which is dependent upon position, rather than a single offset. 

In each pointing of each field we first identified matching objects between DEEP2 and SDSS. This was done by selecting each DEEP2 object and searching for SDSS objects within a given search radius, and in cases where multiple matches are found, the closest object is selected as the match. This general matching procedure was used for all catalog matching in this paper. For all other matching procedures described in this paper we used a search radius of $0.75\arcsec$, as that is approximately the resolution in the DEEP2 survey. However, for the astrometric corrections, we used a larger initial search radius of $1\arcsec$ to allow for systematic errors. For every matched pair of DEEP2 and SDSS objects, we calculated $\ra-\ra_{\mathrm{SDSS}}$ and $\dec-\dec_{\mathrm{SDSS}}$ on a grid by binning in RA and dec and calculating the median difference between DEEP2 and SDSS astrometry in each bin.  For any bins where these differences are poorly constrained, i.e. too few objects to compute a median or with the error in the bin $\gtrsim 0.5\arcsec$ (of order the typical correction factor), we instead use values interpolated from adjoining bins. We then smoothed the gridded offsets to obtain the required corrections to be applied to the original RA and dec values in each pointing to bring them onto the SDSS reference frame. The correction factors for each object were calculated by interpolating on the smoothed grid of values; the results were subtracted from the original positions to yield SDSS-equivalent positions. The refined astrometry was then used to re-match catalogs using our standard 0.75'' search radius.  These corrections resulted in a significant increase in the number of matches found between the two catalogs, ranging from $\sim40-60$ more matches in the shallow SDSS pointings, up to thousands of matches in some of the deeper fields. We investigated iterative refinement of the corrections using this closer match radius beyond the first iteration, but the results did not show significant improvement.

\begin{figure*}[t]
\centering
$\begin{array}{@{\hspace{0in}}c@{\hspace{.2in}}c}
\includegraphics[totalheight=3in]{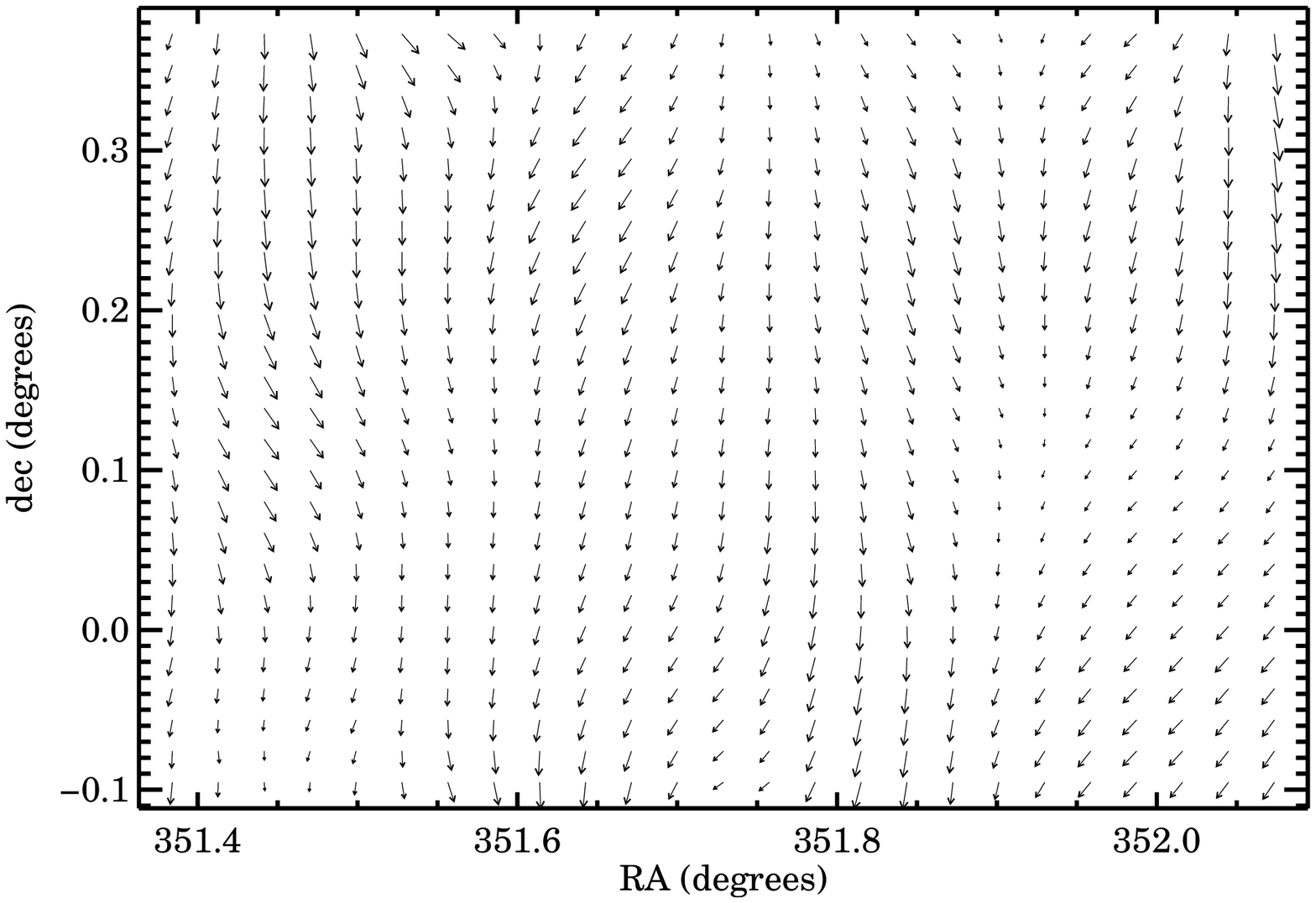}  &
\includegraphics[totalheight=3in]{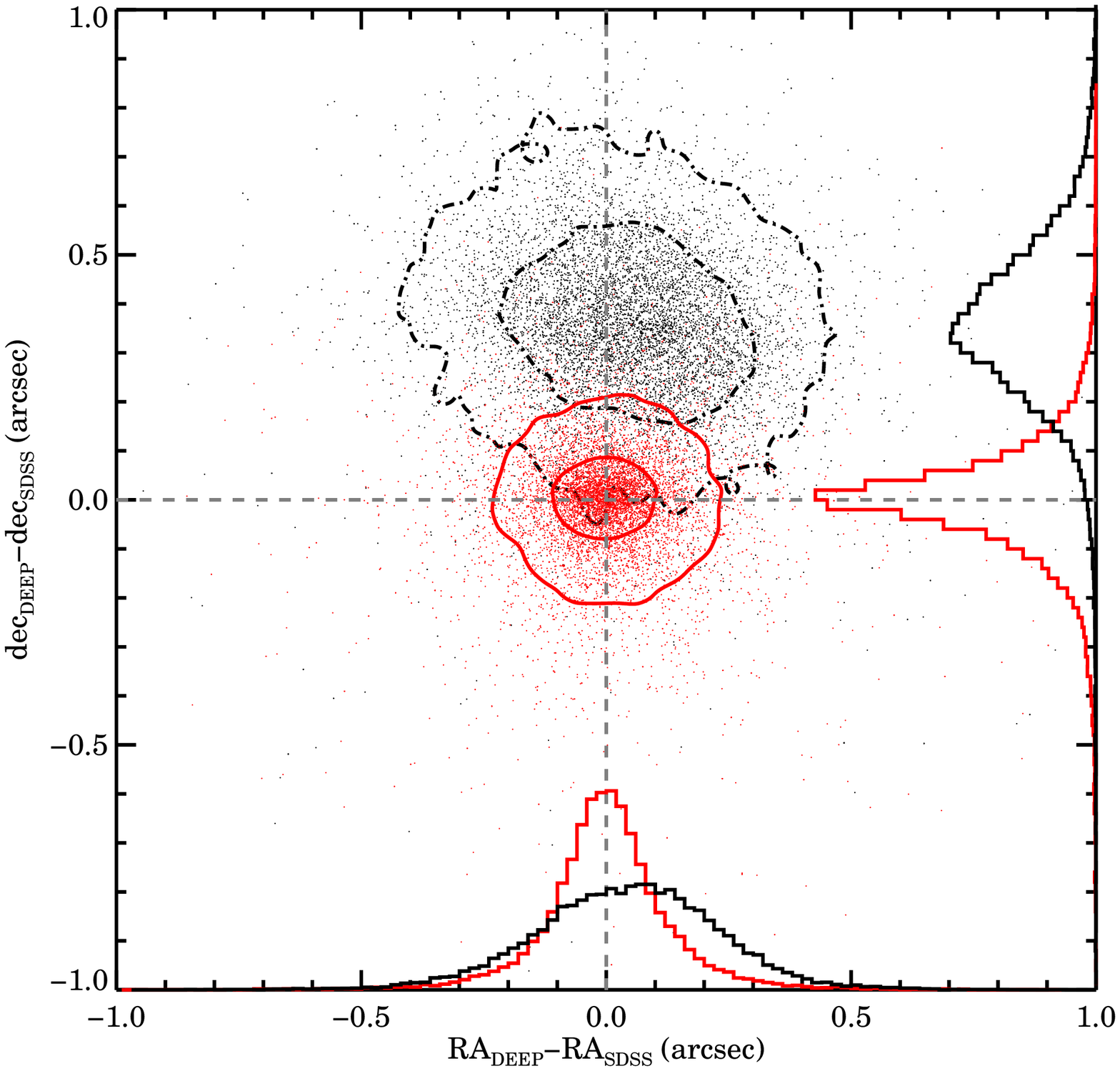} \\
\end{array}$
\caption{The left panel is an arrow plot showing the size and directions of the astrometric corrections applied in pointing 31, where a $0.03^{\circ}$-long arrow indicates a $1\arcsec$ difference. Both the size and direction vary significantly over the field, making that depend on position on small scales necessary. The right panel shows the difference between the DEEP2 and SDSS astrometry for matches in the same pointing, both before (black) and after (red) the correction, where the contour lines correspond to 32\% and 5\% of the peak density, respectively. The projected distributions of each residual are shown on the bottom and right side of the plot, with all histograms normalized to have the same integral. The points show a random subset of all matches, while the contour lines and histograms were constructed using the full set of matches. There is a significant improvement in both the bias and spread after correction for both RA and declination; these differences are quantified for all pointings in Table \ref{tab:astromcorr}.}
\label{fig:radec}
\end{figure*}

Ideally, we would like to perform corrections which can vary on very small scales, in order to capture all possible structure in the astrometric offsets.  However, that would cause only a few objects to be used to determine the correction at any given position, yielding noisy results.  We therefore must adopt a grid scale which balances these two needs.  In order to determine how finely we should bin in RA and dec in order to accurately describe the real deviations at a given position without excessive noise due to using only a small set of objects, we investigated how varying the number of bins we divided the pointing area into affected the rms variation in $\ra-\ra_{\mathrm{SDSS}}-\Delta_{\ra}$ and $\dec-\dec_{\mathrm{SDSS}}-\Delta_{\dec}$ for all matches, where $\Delta$ is the correction factor described above. We repeated the calculation while increasing the number of bins, and in each case smoothed the grid by performing a boxcar average over a width of 5 bins. We started by dividing the pointing area up into 10 bins, and found that in all pointings, the rms variation decreased significantly until reaching around 40-60 bins, where it leveled off.  For all astrometric corrections we set the number of bins equal to 50 (corresponding to $\sim35-50\arcsec$ per bin), as this provided the best balance between fidelity of reconstruction and noise.

As an example, the left panel of Figure \ref{fig:radec} shows the astrometric corrections determined for pointing 31 in DEEP2 Field 3. Table \ref{tab:astromcorr} describes the improvement in astrometry for both CFHTLS-Wide and DEEP2 catalogs resulting from this process.  We list the median and robust standard deviation of $\ra_{SDSS}-\ra$ and $\dec_{SDSS}-\dec$, both before and after corrections are applied.  All standard deviations quoted in this table are derived using a robust estimator, which utilizes the median absolute deviation as an initial estimate and then weights points using Tukey's biweight \citep{1983ured.book.....H}. In every case, there are large improvements in the agreement with SDSS astrometry; the standard deviation of the residuals is dominated by measurement errors, not systematics.  This can be seen in the right panel of Figure \ref{fig:radec} which shows a plot of the astrometric residuals for each calibration object in pointing 31, as well as histograms of their projected distributions in right ascension and declination,  both before and after the correction. The improvement is much greater for DEEP2, but still detectable for the CFHTLS-Wide astrometry.  We utilize the SDSS-reference-frame astrometry for both DEEP2 and CFHTLS in matching objects for the remainder of this paper.

\section{Supplemental Photometric Information for DEEP2 }
\label{sec:deep2}
\subsection{DEEP2 Field 1}
\label{sec:deep2f1}
In DEEP2 Field 1 we provide catalogs for DEEP2 pointings 11, 12, 13 and 14. The $BRI$ photometry for pointings 11, 12, and 13 are taken directly from the DEEP2 catalogs described in \S\ref{sec:datasets}.  These measurements are identical to those provided in DEEP2 Data Release 4 \citep{Newman}. The DEEP2 $BRI$ photometry in pointing 14 had both inferior depth and calibration quality to that obtained in other survey fields due to poor observing conditions when the data were taken.  As a result, a purely $R$-selected sample was targeted in that region \citep{Newman}. However, due to the wide range of multiwavelength data covering that area \citep{2007ApJ...660L...1D}, the redshifts obtained there are quite valuable, and it is desirable to have as uniform a photometric sample as possible.  We have therefore developed improved $BRI$ photometry for the problematic region by using  CFHTLS-Wide $ugriz$ photometry to predict DEEP2 $BRI$. We do this using transformations determined from data in DEEP2 pointings 11 and 12, as described below. We did not use pointing 13 because we found for the DEEP2 data the stellar locus in the color-color relation for that pointing to be not as well determined. In this catalog we also provide the $ugriz$ photometry for all DEEP2 sources that have a matching object (determined as described in \S \ref{sec:astrom}) within either CFHTLS-Wide or Deep. More details of how the $ugriz$ photometry is assigned in described in \S\ref{sec:datatables}.

\begin{figure*}[t]
\centering
\includegraphics[totalheight=4in]{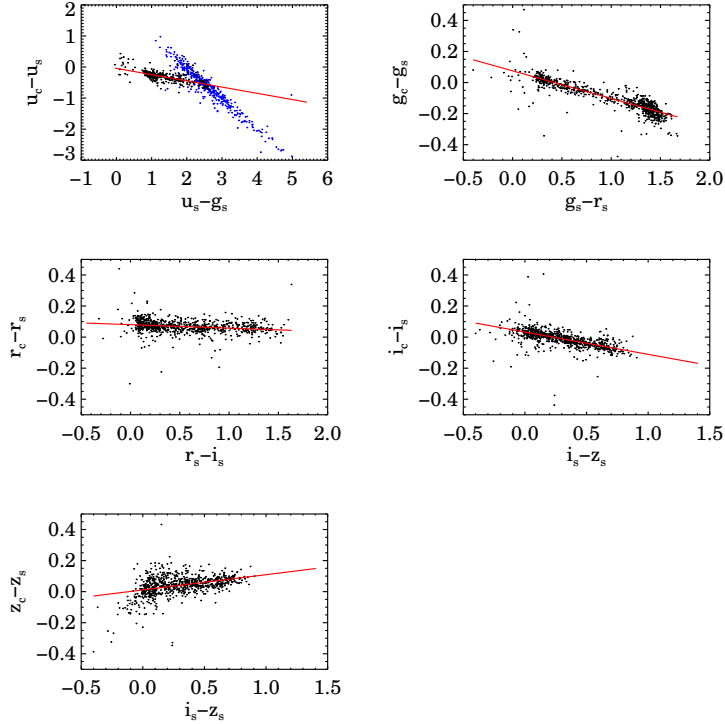}
\caption{Plots of the difference between CFHTLS-Wide and SDSS magnitudes difference as a function of SDSS color term for each $ugriz$ band, utilizing objects identified as stars in SDSS with $18 < r < 20$ that overlap CFHTLS-Wide pointing W3-1-3. The red lines are the linear fits whose coefficients are listed in Table \ref{tab:mcoeff}. In the top left plot we see that there are points that scatter along a second diagonal that does not follow the linear fit. This is due to the large $u$-band measurement errors for objects faint in $u$ in SDSS. We performed a magnitude cut ($u_s<22$) for objects used in the linear fit for this band so that the objects in the second diagonal would not influence the fit, as described in \S\ref{sec:cfhtcal}. Blue points are objects that were not used in the fit.}
\label{fig:calfit}
\end{figure*}

\subsubsection{Improved photometric zero point calibration for CFHTLS data}
\label{sec:cfhtcal}
The CFHTLS-Wide photometry overlapping DEEP2 Field 1 proved to have systematic zero point errors that varied amongst the individual MegaCam pointings.  Hence, it was necessary to recalibrate each CFHTLS-Wide pointing overlapping with DEEP Field 1 in order to provide a uniform dataset. We found that the zero point errors in each band (assessed by comparison to SDSS) varied significantly from pointing to pointing.  The typical offset for a pointing ranged in magnitude from $\sim0.01-0.13$ with typical scatter within a pointing of $\sim 2-4\times10^{-2}$, except for in the $u$-band where the scatter was significantly larger ($\sim0.2$). These calculations are described in detail below. There are seven CFHTLS-Wide MegaCam pointings that overlap the four CFHT 12K pointings in DEEP Field 1: W3-1-2, W3-1-3, W3+0-1, W3+0-2, W3+0-3, W3+1-1, W3+1-2. We calibrated each pointing using objects identified as stars in SDSS DR9 data  with $18 < r < 20$. For each of these stars we determined if there is a match in CFHTLS-Wide by searching for objects within the normal 0.75" search radius, finding an average of 737 matches per pointing.  After finding matches in each catalog we then calculated a linear fit to the magnitude difference between the two bands:
\begin{eqnarray}
\label{eqn:colorfit1}
u_c - u_s &=& a_{0,u} + a_{1,u}(u_s - g_s) \\
g_c - g_s &=& a_{0,g} + a_{1,g}(g_s - r_s) \\
r_c - r_s &=& a_{0,r} + a_{1,r}(r_s - i_s) \\
i_c - i_s &=& a_{0,i} + a_{1,i}(i_s - z_s) \\
\label{eqn:colorfit2}
z_c - z_s &=& a_{0,z} + a_{1,z}(i_s - z_s)
\end{eqnarray}
where the '$c$' subscript denotes CFHTLS-Wide photometry, the '$s$' subscript denotes SDSS DR9 photometry, and $a_0$ and $a_1$ indicate the constant term and slope parameters from the fit. This regression, as well as all other fits done in this paper, were done using the IDL procedure POLY\_ITER from the SDSS idlutils library, which is an iterative fitting procedure that uses outlier rejection.  Adding quadratic color terms did not significantly improve the fits.  Figure \ref{fig:calfit} plots the relations in equations \ref{eqn:colorfit1}-\ref{eqn:colorfit2} as well as the linear fits for pointing W3-1-3. For the $u$-band relation, we used only objects with $u_s<22$ for the fit since $u$-band measurements in SDSS are extremely noisy fainter than this limit, as is evident in the first panel of Figure \ref{fig:calfit}. On average this cut eliminated $\sim57\%$ of the objects from the sample used for fitting. The values for $a_0$ and $a_1$ from the overall fits for each band are listed in Table \ref{tab:mcoeff}. We estimated the uncertainties in these parameters by bootstrapping and found for $a_0$ the errors in the $griz$ bands are $\sim2-6\times10^{-3}$ and $\sim1-3\times10^{-2}$ in the $u$-band. For $a_1$ the uncertainties in the $gri$ bands are $\sim2-6\times10^{-3}$ and $\sim10^{-2}$ and the $u$ and $z$ bands.

If both the CFHT and SDSS photometry were on the AB filter system in their native passbands, we would expect there to be no difference between the magnitudes measured in the two systems for an object with the AB defining spectrum ($F_{\nu}=3.631\times10^{-20}\ \mathrm{erg\ s}^{-1} \mathrm{Hz}^{-1} \mathrm{cm}^{-2}$), which should have the same magnitude in all bands and hence zero color.  Therefore, if $a_0$ is non-zero, $u_c$ and $u_s$ cannot both be on the AB system.  Although SDSS magnitudes are not quite AB, they are very close \citep{1996AJ....111.1748F,2002AJ....123..485S}, and hence we can use the $a_0$ values to determine how the zero points of the CFHTLS-Wide photometry must be changed to place them on a uniform AB system.  In principle, we could perform this fit over small ranges in right ascension and declination to determine the spatial variation in CFHTLS zero point errors; however, in practice SDSS stars are too sparse to measure $a_1$ robustly in small bins of position on the sky.

Instead, we adopt the strategy of using a fixed $a_1$ value for each pointing and determining variation only in $a_0$.   We use a pointing's $a_1$ value for a given band (specified by the above relations) to calculate the quantities
\begin{eqnarray}
\label{eqn:deltaeqn1}
\Delta_u &=& (u_c - u_s) - a_{1,u}(u_s - g_s) \\
\Delta_g &=& (g_c - g_s) - a_{1,g}(g_s - r_s) \\
\Delta_r &=& (r_c - r_s) - a_{1,r}(r_s - i_s) \\
\Delta_i &=& (i_c - i_s) - a_{1,i}(i_s - z_s) \\
\label{eqn:deltaeqn5}
\Delta_z &=& (z_c - z_s) - a_{1,z}(i_s - z_s)
\end{eqnarray}
for each object.  By the same argument given above, if $m_c$ and $m_s$ are AB magnitudes, its offset $\Delta_m$ should be zero everywhere (modulo measurement errors).  For all bands except the $u$-band, we construct a two dimensional map of $\Delta_m$ in RA and dec for each pointing. We expect $\Delta_m$ to vary slowly across the field, and so we determine the map by fitting a 2-D second order polynomial, i.e.
\begin{align}
\label{eqn:deltafit}
\Delta_m(\ra,\dec) =& b_{0,m}+b_{1,m}(\ra)+b_{2,m}(\ra)^2+b_{3,m}(\ra)(\dec) \nonumber \\
                                &+b_{4,m}(\dec)+b_{5,m}(\dec)^2.
\end{align}

The $b$ coefficients are calculated separately for each CFHTLS-Wide pointing and for each passband ($g/r/i/z$). We can then obtain AB-calibrated CFHT photometry in a given band, $m_c'$, by setting $m_c' = m_c - \Delta_m(\ra,\dec)$. For the $u$-band, we obtained better results by calculating a mean $\Delta_u$ in each pointing to obtain $u_c' = u_c - \langle\Delta_u\rangle$, rather than fitting a 2-D polynomial over RA and dec. This was most likely due to noise in the $u$-band measurement affecting the fit. We have used the robust Hodges-Lehmann estimator of the mean to calculate $\langle\Delta_u\rangle$ in each pointing.

We also found it necessary to recalibrate the CFHTLS-Deep photometry in order for it to have consistent zero points with the refined CFHTLS-Wide photometry. We performed this calibration by applying the same techniques used for the Wide survey $u$-band data; i.e., we employ a constant zero point offset, $m_c' = m_c - \langle\Delta_m\rangle$. We adopt this method to make it simple to transform back to the original CFHTLS-Deep photometry, facilitating the use of our catalog to calibrate photo-$z$'s for all of the CFHTLS-Deep fields. We again used the robust Hodges-Lehmann estimator of the mean to calculate $\langle\Delta_m\rangle$. For each band the correction is $\langle\Delta_u\rangle$=-0.01941, $\langle\Delta_g\rangle$=0.07374, $\langle\Delta_r\rangle$= 0.03056, $\langle\Delta_i\rangle$=0.04441, and $\langle\Delta_z\rangle$=0.03282.  The values for $a_0$ and $a_1$ (Equations \ref{eqn:colorfit1}-\ref{eqn:colorfit2}) calculated for CFHTLS-Deep photometry are listed in Table \ref{tab:mcoeff}.

\begin{figure}[t]
\centering
\includegraphics[totalheight=3.4in]{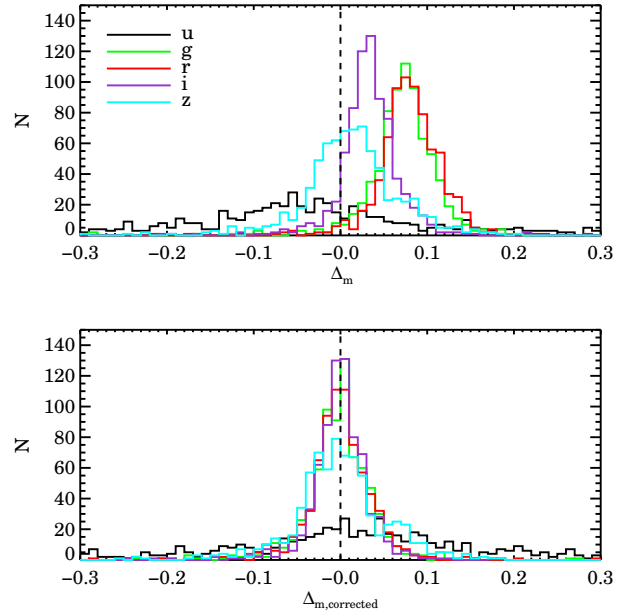}
\caption{The distributions of the zero-point offsets for the CFHTLS-Wide photometry in pointing W3-1-3 relative to SDSS DR9 ($\Delta_m$ in equations \ref{eqn:deltaeqn1}-\ref{eqn:deltaeqn5}) for the bright stars ($18<r<20$) in each band before and after the improved calibration. After the corrections to the $ugriz$ photometry, the systematic offsets in each band are removed. This improvement is shown quantitatively for all pointings in Table \ref{tab:zeropt}.}
\label{fig:zeropt}
\end{figure}

These corrections have been applied to the CFHTLS-Wide and Deep $ugriz$ photometry for all objects in this catalog. Table \ref{tab:zeropt} shows the improvement in the zero-point offset estimate for each pointing by showing the median and standard deviation of this offset amongst all SDSS reference stars before and after this calibration, and Figure \ref{fig:zeropt} shows the distribution of these offsets for all $ugriz$ bands in pointing W3-1-3 both before and after the calibration.  All standard deviations quoted in the table are calculated using the robust estimator described in \S\ref{sec:astrom}.  Median offsets become negligible after correction; zero point errors that in some cases approached 0.2 mag are $\lesssim 0.01$ after correction.  The standard deviation is dominated by random uncertainties, but still is reduced in all but one case, indicating that our spatially-varying zero point correction has improved the match between CFHTLS-Wide and SDSS photometry compared to a uniform offset.

The $a_1$ coefficients calculated from the linear fits in Equations \ref{eqn:colorfit1}-\ref{eqn:colorfit2} can also be used to transform between the CFHTLS and SDSS photometry systems. For example, due to how we have defined the zero point offset for our new calibrated CFHTLS photometry, the transformation for the $u$-band is defined as $u_c' = u_s + \langle a_{1,u}\rangle(u_s - g_s)$, where we have calculated $\langle a_{1,u}\rangle$ from the average $a_{1,u}$ values over all seven CFHTLS-Wide pointings and the single CFHTLS-Deep pointing. The color terms used in the transformations for all other bands can be determined from Equations \ref{eqn:colorfit1}-\ref{eqn:colorfit2}, and the values of $\langle a_1\rangle$ for each band are listed in Table \ref{tab:mcoeff}. This transformation can be applied to bring SDSS photometry into the same filter system as CFHTLS. In addition, solving for the SDSS magnitude in the above equation allows for the transformation of the CFHTLS photometry from this catalog into the same system as SDSS. Either transformation can bring the entire catalog into the same $ugriz$ system for photo-$z$ tests.

\begin{figure*}[t]
\centering
\includegraphics[totalheight=4in]{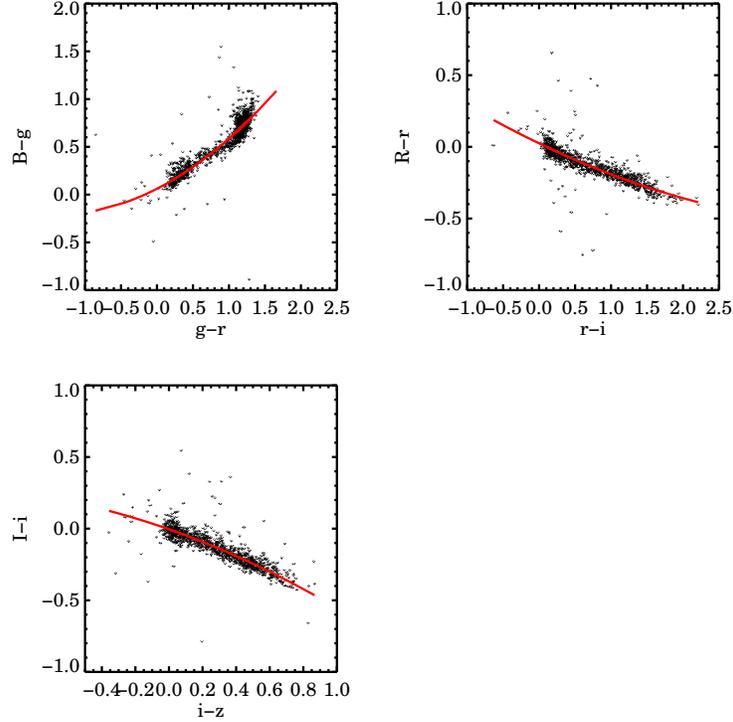}
\caption{Plots of the difference between DEEP2 and CFHTLS-Wide magnitudes as a function of  CFHTLS color for each $BRI$ band, using objects identified as stars in DEEP2 pointings 11 and 12 with $18.2 < R < 21$. The red lines are the quadratic fits whose coefficients are listed in Table \ref{tab:bricoeff}.}
\label{fig:deepfit}
\end{figure*}

\subsubsection{Predicting photometry of DEEP pointing 14}
\label{sec:predphot}
Due to the inferior photometry in DEEP2 pointing 14, we used the CFHTLS-Wide $ugriz$ photometry for objects in pointings 11 and 12 to predict the $BRI$ photometry in DEEP pointing 14. 
To determine the transformation between the two systems, we chose sources identified as stars in the DEEP2 catalog with $18.2 < R < 21$. This range was selected in order to obtain a sample of bright stars which are also above the saturation limit of the DEEP2 survey. 

The $BRI$ photometry in the DEEP2 catalogs have been corrected for Galactic dust extinction \citep{1998ApJ...500..525S}. However in the CFHTLS-Wide catalog, magnitudes have not been adjusted for this.  Hence, before determining color transformations, we removed the extinction correction from the DEEP2 $BRI$ photometry, using the same  \citet{1998ApJ...500..525S} reddening estimates (SFD\_EBV) and $R$ values that were employed to make the original DEEP2 catalogs.

\begin{figure}[t]
\centering
\includegraphics[totalheight=3.4in]{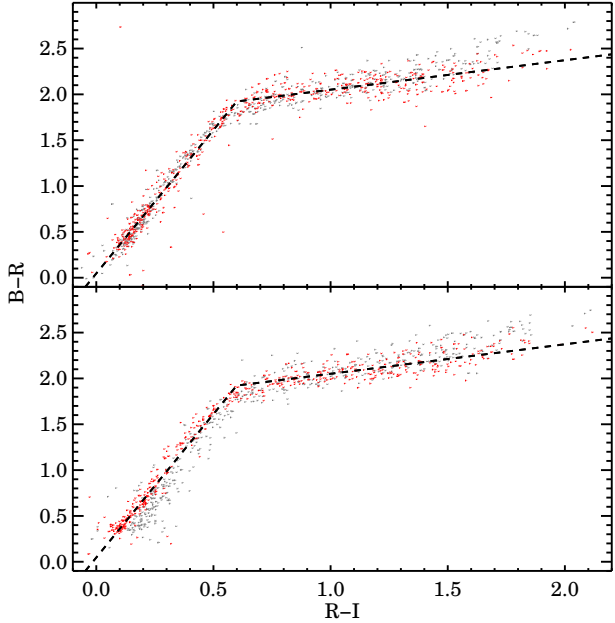}
\caption{Color-color plots of the stars with $18.2 < R < 21$ that were used to determine the $ugriz$ (CFHTLS-Wide) to $BRI$ (DEEP2) transformation described in \S\ref{sec:predphot}. The top panel shows the stellar locus for pointing 11 and the bottom for pointing 14, and the dashed lines are the same in each plot. The gray points are the colors straight from the public DEEP2 catalogs, and the red points are the colors after the transformation. The stellar locus in pointing 11 is relatively unaffected by the transformation compared to pointing 14. The improved calibration of the pointing 14 photometry is apparent in the greater consistency of the stellar locus for pointing 14 after the transformation to that from pointing 11 (most easily visible by comparing each to the dashed lines).}
\label{fig:deepcolor}
\end{figure}

We matched these sources to CFHTLS-Wide objects again using a 0.75" search radius, and calculated the parameters of the relations
\begin{eqnarray}
\label{eqn:brieqn1}
B-g &=& c_{0,B} + c_{1,B}(g-r) + c_{2,B}(g-r)^2 \\
\label{eqn:brieqn2}
R-r &=& c_{0,R} + c_{1,R}(r-i) + c_{2,R}(r-i)^2 \\
\label{eqn:brieqn3}
I-i &=& c_{0,I} + c_{1,I}(i-z) + c_{2,I}(i-z)^2.
\end{eqnarray}
Figure \ref{fig:deepfit} plots the relations in equations \ref{eqn:brieqn1}-\ref{eqn:brieqn3} as well as the quadratic fit for pointings 11 and 12. We then use these parameters to calculate the predicted photometry for all objects in DEEP pointing 14, including sources identified as galaxies. By plotting the residuals of all objects as a function of $r$-band half light radius as determined in the CFHTLS-Wide catalog, we found there is a contribution from the source size. We represented this contribution with a linear fit to these residuals, which gives the final predicted photometry as
\begin{eqnarray}
\label{eqn:britoteqn1}
B &=& g + c_{0,B} + c_{1,B}(g-r) + c_{2,B}(g-r)^2 + d_{0,B} + d_{1,B}(\mathrm{R}_r) \\
R &=& r + c_{0,R} + c_{1,R}(r-i) + c_{2,R}(r-i)^2  + d_{0,R} + d_{1,R}(\mathrm{R}_r) \\
\label{eqn:britoteqn3}
I &=& i + c_{0,I} + c_{1,I}(i-z) + c_{2,I}(i-z)^2 + d_{0,I} + d_{1,I}(\mathrm{R}_r)
\end{eqnarray}
where $\mathrm{R}_r$ is the $r$-band half light radius (designated as r\_flux\_radius in the catalog). Table \ref{tab:bricoeff} lists all of these coefficients for pointings 11-13 as well as the coefficients calculated from combining pointings 11 and 12. Errors in this predicted photometry were calculated using simple propagation of errors using the errors in $g,r,i,$ and $z$ from CFHTLS-Wide. In order to maintain consistency with DEEP2 photometry in other fields, we then apply a correction for extinction in the same manner as for the other DEEP2 magnitudes. 

Figure \ref{fig:deepcolor} shows color-color plots for the bright stars in pointings 11 and 14 that were used to determine the $griz$ to $BRI$ transformation described above, both before and after applying the transformation. We see that the stellar locus in pointing 11 is relatively unaffected by the transformation compared to pointing 14. We also see the improved calibration of the pointing 14 photometry in that the locus in 14 is tighter and more consistent with the locus in 11 after the transformation. We note that although we applied the transformation to obtain predicted $BRI$ photometry in pointing 11 for this plot, in the final catalog the transformation is only applied to pointing 14.

\subsection{DEEP2 Fields 2, 3 and 4}
\label{sec:deep2f234}
We are also providing catalogs with improved astrometry (cf. \S \ref{sec:astrom}) and $ugriz$ photometry added for objects from DEEP2 Field 2 (pointings 21 \& 22), Field 3 (pointings 31,32, \& 33) and DEEP Field 4 (pointings 41, 42, \& 43). In each case, the $BRI$ photometry is taken directly from the DEEP2 catalogs described in \S\ref{sec:datasets}, while the $ugriz$ photometry is determined from matching sources in SDSS, using the procedure described in \S\ref{sec:astrom}.  In Field 2 we use SDSS photometry from the DR9 data release.  Since Fields 3 and 4 overlap with Stripe 82, we use the deeper photometry from the Stripe82 database (cf. \S\ref{sec:datasets}).

\begin{figure*}[t]
\centering
\includegraphics[totalheight=5.9in]{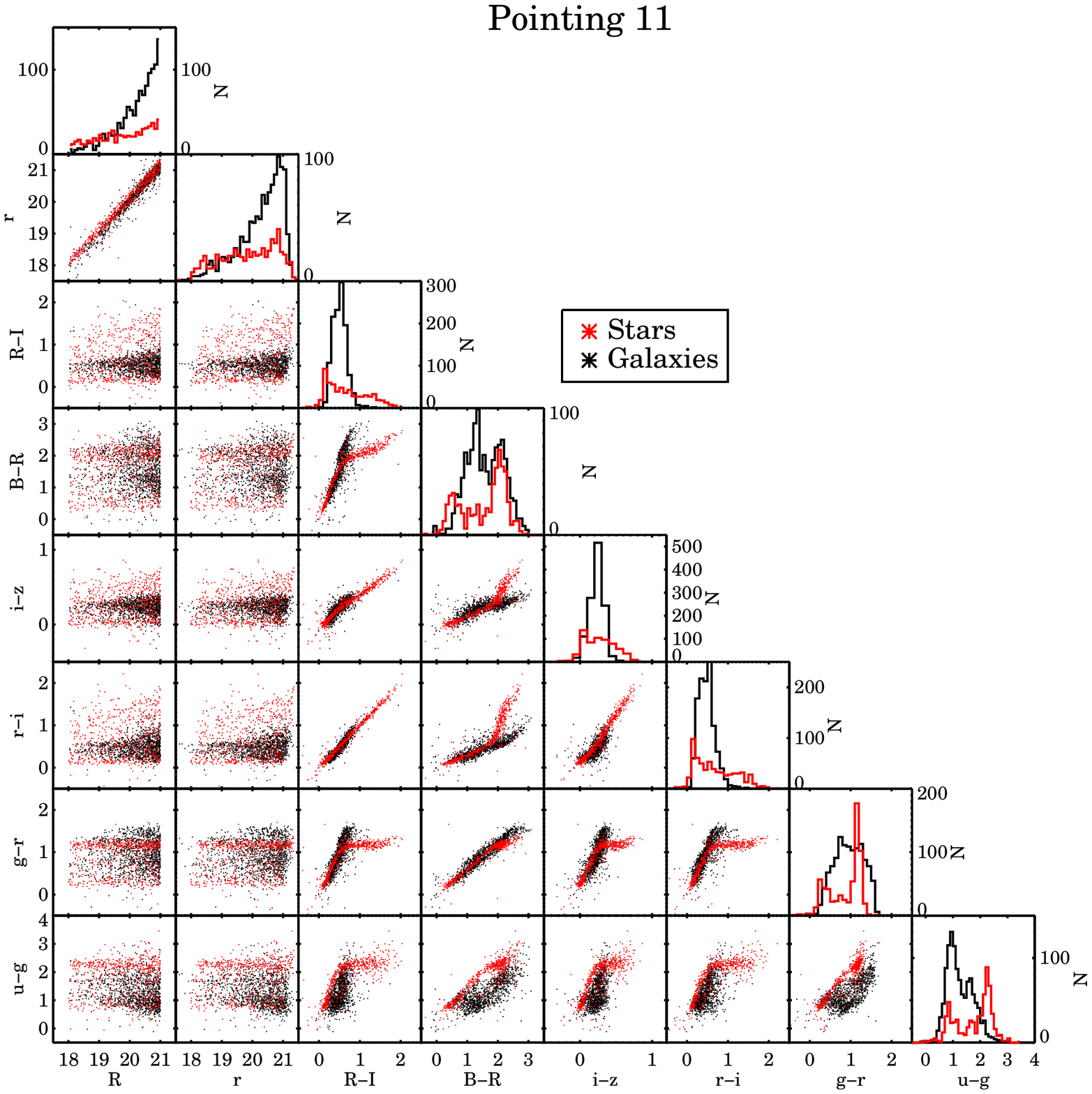}
\caption{Plots of the relations between various photometric quantities for bright stars (red) and galaxies (black) with $18 < R < 21$ that have photometry in both DEEP2 pointing 11 and in CFHTLS-Wide. Histograms of each quantity are shown on the diagonal.}
\label{fig:color}
\end{figure*}

\section{Data Tables}
\label{sec:datatables}

Below we describe the columns that are included for each object in our new FITS BINTABLE-format files, as well as a brief description of each quantity. We have created one set of new catalogs that are parallel in content to each of the existing {\it pcat} photometric catalogs, as well as a single new catalog that parallels the {\it zcat} redshift catalog, and hence contains only objects for which DEEP2 obtained a spectrum. All of these columns appear in both catalogs. Further details of those columns that have been taken directly from other catalogs can be found in \citet{Newman}, \citet{2012AJ....143...38G}, \citet{2009ApJS..182..543A}, and \citet{2012ApJS..203...21A}. For each object, any column where data is not available is given a value of -99. The object properties included in the catalog are:

\begin{itemize}
 \item OBJNO - a unique 8-digit object identification number, taken from the {\it pcat} photometric catalog.  The first digit of OBJNO indicates the DEEP2 field an object is drawn from, and the second object indicates pointing number (e.g., objects in DEEP2 pointing 14 will have object numbers beginning with 14).
 \item RA\subscript{DEEP}, dec\subscript{DEEP} - Right ascension and declination in degrees from the DEEP2 catalogs, including the astrometric correction described in \S\ref{sec:astrom}.  These positions will therefore differ from those in the original {\it pcat} catalogs.
 \item RA\subscript{SDSS}, dec\subscript{SDSS} - Right ascension and declination in degrees from either CFHTLS-Wide (DEEP2 Field 1) or SDSS (DEEP2 Fields 2-4) for all {\it pcat} objects that have a match in either catalog. The CFHTLS-Wide astrometry in Field 1 has been corrected as described in \S\ref{sec:astrom}.
 \item BESTB, BESTR, BESTI - For all pointings except for pointing 14 in Field 1, these are CFHT 12K $BRI$ magnitudes taken directly from the DEEP2 {\it pcat} catalogs. Photometry in pointing 14 is predicted using the methods described in \S\ref{sec:predphot} for objects that have a match with CFHTLS-Wide. Objects without a match are assigned no $BRI$ values in pointing 14.
 \item BESTBERR, BESTRERR, BESTIERR - errors in the $BRI$ photometry taken directly from the DEEP2 catalogs for all pointings except for pointing 14 in Field 1. Those error estimates include sky noise only.  Errors for pointing 14 were calculated using simple propagation of errors from the errors in CFHTLS-Wide photometry.
 \item U, G, R, I, Z - $ugriz$ magnitudes taken either from CFHTLS-Wide (DEEP2 Field 1) or SDSS (DEEP2 Fields 2-4) for all {\it pcat} objects that have a  match in either catalog. The CFHTLS photometry used was the Kron-like elliptical aperture magnitude designated as MAG\_AUTO in the unified CFHTLS catalogs. In our new {\it zcat} catalog, if photometry is available for an object in Field 1 from the CFHTLS-Deep survey, that is used; otherwise the magnitudes from the CFHTLS-Wide survey are used. In our new {\it pcat} catalogs, CFHTLS-Wide is used for all photometry. SDSS magnitudes in Fields 2-4 are the model magnitudes taken from either DR9 (Field 2) or the coadded Stripe 82 database (Fields 3 \& 4).
 \item UERR, GERR, RERR, IERR, ZERR - errors in the $ugriz$ magnitudes taken directly from either CFHTLS-Wide or Deep (for DEEP2 Field 1), or from SDSS (DEEP2 Fields 2-4), for all {\it pcat} objects that have a  match in either catalog.
 \item PGAL - probability of the object being a galaxy based on the $R$-band image and $BRI$ color. For the calculations in this paper, any object with $p_{gal}<0.2$ was treated as a star, following the standard in \citet{Newman}.
 \item RG - Gaussian radius of a circular 2-d Gaussian fit to the R-band image, in units of 0.207\arcsec CFHT 12K pixels.
 \item BADFLAG - quantity describing the quality of the $BRI$ photometry measurement. A badflag value of zero designates a measurement with no known systematic issues (e.g. saturation, overlapping with bleed trails, etc.) in any bands (http://deep.berkeley.edu/ DR1/photo.primer.html).
 \item ZHELIO - heliocentric reference-frame redshift taken from the {\it zcat} catalogs.
 \item ZHELIO\_ERR - error in the redshift measurement taken from the {\it zcat} catalogs.
 \item ZQUALITY - redshift quality code, Q, where $Q = 3$ or 4 indicates a reliable galaxy redshifts, and $Q=-1$ indicates a reliable star identification.
 \item SFD\_EBV - Galactic reddening $E(B-V)$ from \citet{1998ApJ...500..525S}.  DEEP2 $BRI$ photometry has been corrected for this amount of reddening.
 \item SOURCE - string describing the source of the photometry for each object, where the first catalog listed is the source of the $BRI$ photometry and the second is the source of the $ugriz$ photometry. For DEEP2 pointings 11-13 the the source tag is either DEEP-CFHTLSW or DEEP-CFHTLSD (Wide or Deep), and for pointing 14 it is just CFHTLSW since the $BRI$ is predicted from CFHTLS-Wide. In DEEP2 Field 2 the source tag is DEEP-SDSS and for Fields 3 and 4 the source tag is DEEP-SDSS82, designating that the $ugriz$ photometry comes from the deeper Stripe82 database. For all objects lacking a match in other catalogs the source tag is just DEEP.
\end{itemize}
Table 5 shows examples of the catalog data for nine objects in pointing 11; three objects with no matches between DEEP2 and CFHTLS, three objects with matches but no redshifts, and three objects with matches and redshifts.

\section{Summary and Conclusions}
\label{sec:conclusion}
In this paper we have presented the details of improved photometric catalogs for the DEEP2 Galaxy Redshift Survey constructed by combining data from three different projects: DEEP2 itself, the CFHT Legacy Survey, and SDSS. To further this purpose, we have used positions from SDSS to improve the astrometry for both DEEP2 and CFHTLS-Wide catalogs, and photometry for SDSS stars to improve the magnitude zero points in the CFHTLS-Wide data.

We then employed data from CFHTLS and SDSS to assign $ugriz$ photometry to DEEP2 objects by matching sky positions between each catalog.  In DEEP2 Field 1 we matched to the CFHTLS-Wide or Deep, in Field 2 we matched to SDSS DR9, and in Fields 3 and 4 we used the deeper SDSS Stripe 82 database. For objects in DEEP2 pointing 14 that had a counterpart in CFHTLS-Wide, we replaced the poorer-than-standard $BRI$ photometry with predicted values calculated using the transformations between the $BRI$ and $ugriz$ photometry measured in DEEP2 pointings 11 and 12. 

In each of the four pointings of DEEP2 Field 1 there are an average of $\sim40,000$ matches with CFHTLS. Figure \ref{fig:color} shows the relations between various photometric quantities for bright stars and galaxies ($18 < R < 21$) that have matches between DEEP2 and CFHTLS in pointing 11. For pointing 14 where we have predicted the $BRI$ photometry from the CFHTLS photometry, the equivalent figure looks qualitatively similar with the exception of any plot that relates DEEP2 color (i.e. $B-R$ or $R-I$) to the CFHTLS colors used to calculate the transformations in equations \ref{eqn:britoteqn1}-\ref{eqn:britoteqn3}. These relations look necessarily tighter since the $BRI$ photometry was calculated using a fit to these color relations. The $r$ vs $R$ relation looks significantly tighter as well for similar reasons. The total number of objects over all four pointings that have both $ugriz$ photometry and spectroscopic measurements is 16,584;  11,897 of those have  high quality redshift measurements (zquality $\ge 3$). 

In the two pointings of DEEP2 Field 2 the average number of matches is only $\sim7,700$ per CFHT 12K pointing, due to the shallowness of the SDSS DR9 dataset which overlaps the field. The total number of objects with both $ugriz$ photometry and redshifts in Field 2 is 968, with 751 having high quality redshifts. The three pointings of Field 3, where deeper Stripe 82 photometry is available, average $\sim 19,600$ matches between the {\it pcat} catalogs and SDSS. The total number of objects in Field 3 with $ugriz$ photometry and redshifts is 9691, with 6947 having high quality redshift measurements. Field 4 also overlaps SDSS Stripe 82; it includes three CFHT 12K pointings with an average of $\sim22,200$ matches each, yielding 9445 objects with $ugriz$ photometry and redshifts, 6987 of which have secure redshifts. 

In this paper we have paired the spectroscopic redshift measurements from the DEEP2 Survey with the $ugriz$ photometry of CFHTLS an SDSS, making this catalog a valuable resource for the future as an excellent testbed for photo-$z$ studies. These catalogs would be useful to future surveys such as LSST and DES for testing photo-$z$ algorithms as well as the calibration of photo-$z$'s. There are few public catalogs available with this number of objects with full $ugriz$ photometry as well as quality redshifts to this depth ($z\sim1.4$). As a comparison, the zCOSMOS data release DR2 is one of the larger current datasets with these characteristics, and it contains $\sim10,000$ objects with $ugriz$ photometry out to $z\sim0.8$, $\sim$6000 of which have secure redshifts \citep{2009ApJS..184..218L}.  We caution readers that the SDSS and CFHTLS $ugriz$ passbands differ, so the combined redshift and photometric catalog presented here should not be treated as a uniform dataset; however, the SOURCE column can be used to divide into separate catalogs with consistent photometric passbands, which can be used separately to test photometric redshift methods. Alternatively, the CFHTLS or SDSS photometry can be transformed as described in \S\ref{sec:cfhtcal}, bringing the entire catalog into the same $ugriz$ system. For future work we plan to use Y-band imaging from the Subaru Telescope Suprime-Cam that is currently being taken in the Extended Groth Strip to supplement the ugriz $ugriz$ photometry, which will provide an LSST-like photometric dataset for testing photo-$z$ algorithms. 

The extended DEEP2 catalogs described in this paper are publicly available and can be downloaded at http://deep.ps.uci.edu/DR4/photo.extended.html.

We thank the referee for helpful comments. This work was supported by the United States Department of Energy Early Career program via grant DE-SC0003960 and by the National Science Foundation via grant AST-0806732.

Funding for the DEEP2 Galaxy Redshift Survey has been provided by NSF grants AST-95-09298, AST-0071048, AST-0507428, and AST-0507483 as well as NASA LTSA grant NNG04GC89G. Some of the data included in the DEEP2 DR4 were obtained at the W. M. Keck Observatory, which is operated as a scientific partnership among the California Institute of Technology, the University of California, and the National Aeornautics and Space Administration. The Observatory was made possible by the generous financial support of the W. M. Keck Foundation. The DEIMOS spectrograph was funded by grants from CARA (Keck Observatory) and UCO/Lick Observatory, a NSF Facilities and Infrastructure grant (ARI92-14621), the Center for Particle Astrophysics, and by gifts from Sun Microsystems and the Quantum Corporation. Primary imaging data for the DEEP2 survey was collected using the Canada France Hawaii Telescope (CFHT), which is operated by the National Research Council (NRC) of Canada, the Institute National des Sciences de l'Univers of the Centre National de la Recherche Scientifique of France, and the University of Hawaii. The authors would also like to recognize and acknowledge the highly significant cultural role and reverence that the summit of Mauna Kea has always had within the indigenous Hawaiian community. We are fortunate to be given the opportunity to conduct observations from this mountain.

    Funding for the SDSS and SDSS-II has been provided by the Alfred P. Sloan Foundation, the Participating Institutions, the National Science Foundation, the U.S. Department of Energy, the National Aeronautics and Space Administration, the Japanese Monbukagakusho, the Max Planck Society, and the Higher Education Funding Council for England. The SDSS Web Site is http://www.sdss.org/. The SDSS is managed by the Astrophysical Research Consortium for the Participating Institutions. The Participating Institutions are the American Museum of Natural History, Astrophysical Institute Potsdam, University of Basel, University of Cambridge, Case Western Reserve University, University of Chicago, Drexel University, Fermilab, the Institute for Advanced Study, the Japan Participation Group, Johns Hopkins University, the Joint Institute for Nuclear Astrophysics, the Kavli Institute for Particle Astrophysics and Cosmology, the Korean Scientist Group, the Chinese Academy of Sciences (LAMOST), Los Alamos National Laboratory, the Max-Planck-Institute for Astronomy (MPIA), the Max-Planck-Institute for Astrophysics (MPA), New Mexico State University, Ohio State University, University of Pittsburgh, University of Portsmouth, Princeton University, the United States Naval Observatory, and the University of Washington.

Funding for SDSS-III has been provided by the Alfred P. Sloan Foundation, the Participating Institutions, the National Science Foundation, and the U.S. Department of Energy Office of Science. The SDSS-III web site is http://www.sdss3.org/. SDSS-III is managed by the Astrophysical Research Consortium for the Participating Institutions of the SDSS-III Collaboration including the University of Arizona, the Brazilian Participation Group, Brookhaven National Laboratory, University of Cambridge, Carnegie Mellon University, University of Florida, the French Participation Group, the German Participation Group, Harvard University, the Instituto de Astrofisica de Canarias, the Michigan State/Notre Dame/JINA Participation Group, Johns Hopkins University, Lawrence Berkeley National Laboratory, Max Planck Institute for Astrophysics, Max Planck Institute for Extraterrestrial Physics, New Mexico State University, New York University, Ohio State University, Pennsylvania State University, University of Portsmouth, Princeton University, the Spanish Participation Group, University of Tokyo, University of Utah, Vanderbilt University, University of Virginia, University of Washington, and Yale University. 

This work is based in part on data products produced at the Canadian Astronomy Data Centre and Terapix as part of the Canada-France-Hawaii Telescope Legacy Survey, a collaborative project of the National Research Council of Canada and the French Centre national de la recherche scientifique.

\bibliography{adsreferences}


\begin{table}[h]
\small
\begin{center}
\begin{tabular}{|c|cc|cc|} \hline
\multirow{2}{*}{Pointing} & \multicolumn{2}{|c|}{$\ra_{SDSS}-\ra_{DEEP}$ (\arcsec)} & \multicolumn{2}{|c|}{$\dec_{SDSS}-\dec_{DEEP}$ (\arcsec)} \\
 & median & $\sigma$ & median & $\sigma$ \\ \hline
\multirow{2}{*}{11} & $-6.20\!\times\!10^{\text{-}1}$ & $ 3.01\!\times\!10^{\text{-}1}$ & $-6.51\!\times\!10^{\text{-}2}$ & $ 1.89\!\times\!10^{\text{-}1}$ \\
  & ${\it -8.17\!\times\!10^{\text{-}4}}$ & ${\it  1.83\!\times\!10^{\text{-}1}}$ & ${\it -6.68\!\times\!10^{\text{-}3}}$ & ${\it  1.07\!\times\!10^{\text{-}1}}$ \\ \hline
\multirow{2}{*}{12} & $-6.42\!\times\!10^{\text{-}1}$ & $ 3.24\!\times\!10^{\text{-}1}$ & $-2.46\!\times\!10^{\text{-}1}$ & $ 2.44\!\times\!10^{\text{-}1}$ \\
  & ${\it -3.77\!\times\!10^{\text{-}3}}$ & ${\it  1.95\!\times\!10^{\text{-}1}}$ & ${\it -3.47\!\times\!10^{\text{-}3}}$ & ${\it  1.14\!\times\!10^{\text{-}1}}$ \\ \hline
\multirow{2}{*}{13} & $-7.04\!\times\!10^{\text{-}1}$ & $ 3.55\!\times\!10^{\text{-}1}$ & $-4.02\!\times\!10^{\text{-}1}$ & $ 1.80\!\times\!10^{\text{-}1}$ \\
  & ${\it -5.41\!\times\!10^{\text{-}3}}$ & ${\it  1.75\!\times\!10^{\text{-}1}}$ & ${\it -5.19\!\times\!10^{\text{-}3}}$ & ${\it  1.02\!\times\!10^{\text{-}1}}$ \\ \hline
\multirow{2}{*}{14} & $-4.32\!\times\!10^{\text{-}1}$ & $ 3.43\!\times\!10^{\text{-}1}$ & $-2.70\!\times\!10^{\text{-}1}$ & $ 1.88\!\times\!10^{\text{-}1}$ \\
  & ${\it -1.86\!\times\!10^{\text{-}3}}$ & ${\it  1.87\!\times\!10^{\text{-}1}}$ & ${\it -2.60\!\times\!10^{\text{-}3}}$ & ${\it  1.03\!\times\!10^{\text{-}1}}$ \\ \hline \hline
\multirow{2}{*}{21} & $-3.18\!\times\!10^{\text{-}1}$ & $ 1.39\!\times\!10^{\text{-}1}$ & $-1.97\!\times\!10^{\text{-}1}$ & $ 9.75\!\times\!10^{\text{-}2}$ \\
  & ${\it  3.27\!\times\!10^{\text{-}5}}$ & ${\it  9.38\!\times\!10^{\text{-}2}}$ & ${\it -1.08\!\times\!10^{\text{-}3}}$ & ${\it  7.67\!\times\!10^{\text{-}2}}$ \\ \hline
\multirow{2}{*}{22} & $-2.65\!\times\!10^{\text{-}1}$ & $ 1.20\!\times\!10^{\text{-}1}$ & $-1.82\!\times\!10^{\text{-}1}$ & $ 9.73\!\times\!10^{\text{-}2}$ \\
  & ${\it -2.84\!\times\!10^{\text{-}3}}$ & ${\it  9.04\!\times\!10^{\text{-}2}}$ & ${\it -2.01\!\times\!10^{\text{-}4}}$ & ${\it  7.29\!\times\!10^{\text{-}2}}$ \\ \hline \hline
\multirow{2}{*}{31} & $-4.61\!\times\!10^{\text{-}2}$ & $ 1.86\!\times\!10^{\text{-}1}$ & $-3.54\!\times\!10^{\text{-}1}$ & $ 1.58\!\times\!10^{\text{-}1}$ \\
  & ${\it -1.08\!\times\!10^{\text{-}4}}$ & ${\it  1.15\!\times\!10^{\text{-}1}}$ & ${\it -5.98\!\times\!10^{\text{-}4}}$ & ${\it  1.03\!\times\!10^{\text{-}1}}$ \\ \hline
\multirow{2}{*}{32} & $-8.83\!\times\!10^{\text{-}2}$ & $ 2.14\!\times\!10^{\text{-}1}$ & $-3.14\!\times\!10^{\text{-}1}$ & $ 1.86\!\times\!10^{\text{-}1}$ \\
  & ${\it  1.30\!\times\!10^{\text{-}3}}$ & ${\it  1.16\!\times\!10^{\text{-}1}}$ & ${\it -9.10\!\times\!10^{\text{-}4}}$ & ${\it  1.07\!\times\!10^{\text{-}1}}$ \\ \hline
\multirow{2}{*}{33} & $ 8.48\!\times\!10^{\text{-}2}$ & $ 1.87\!\times\!10^{\text{-}1}$ & $-5.67\!\times\!10^{\text{-}1}$ & $ 1.53\!\times\!10^{\text{-}1}$ \\
  & ${\it -5.99\!\times\!10^{\text{-}4}}$ & ${\it  1.13\!\times\!10^{\text{-}1}}$ & ${\it -4.95\!\times\!10^{\text{-}4}}$ & ${\it  9.71\!\times\!10^{\text{-}2}}$ \\ \hline \hline
\multirow{2}{*}{41} & $ 1.27\!\times\!10^{\text{-}1}$ & $ 1.71\!\times\!10^{\text{-}1}$ & $-3.37\!\times\!10^{\text{-}1}$ & $ 1.79\!\times\!10^{\text{-}1}$ \\
  & ${\it  5.24\!\times\!10^{\text{-}4}}$ & ${\it  1.07\!\times\!10^{\text{-}1}}$ & ${\it -4.70\!\times\!10^{\text{-}4}}$ & ${\it  1.02\!\times\!10^{\text{-}1}}$ \\ \hline
\multirow{2}{*}{42} & $ 1.02\!\times\!10^{\text{-}1}$ & $ 1.70\!\times\!10^{\text{-}1}$ & $-2.92\!\times\!10^{\text{-}1}$ & $ 1.66\!\times\!10^{\text{-}1}$ \\
  & ${\it -3.72\!\times\!10^{\text{-}4}}$ & ${\it  1.09\!\times\!10^{\text{-}1}}$ & ${\it -3.69\!\times\!10^{\text{-}5}}$ & ${\it  1.02\!\times\!10^{\text{-}1}}$ \\ \hline
\multirow{2}{*}{43} & $ 3.15\!\times\!10^{\text{-}2}$ & $ 1.64\!\times\!10^{\text{-}1}$ & $-3.34\!\times\!10^{\text{-}1}$ & $ 1.72\!\times\!10^{\text{-}1}$ \\
  & ${\it  1.21\!\times\!10^{\text{-}3}}$ & ${\it  1.07\!\times\!10^{\text{-}1}}$ & ${\it -6.59\!\times\!10^{\text{-}4}}$ & ${\it  1.04\!\times\!10^{\text{-}1}}$ \\ \hline \hline
\multirow{2}{*}{Pointing} & \multicolumn{2}{|c|}{$\ra_{SDSS}-\ra_{CFHT}$ (\arcsec)} & \multicolumn{2}{|c|}{$\dec_{SDSS}-\dec_{CFHT}$ (\arcsec)} \\
 & median & $\sigma$ & median & $\sigma$ \\ \hline
\multirow{2}{*}{11} & $ 1.04\!\times\!10^{\text{-}3}$ & $ 2.10\!\times\!10^{\text{-}1}$ & $ 1.42\!\times\!10^{\text{-}3}$ & $ 1.30\!\times\!10^{\text{-}1}$ \\
  & ${\it  3.40\!\times\!10^{\text{-}4}}$ & ${\it  1.99\!\times\!10^{\text{-}1}}$ & ${\it -4.17\!\times\!10^{\text{-}3}}$ & ${\it  1.16\!\times\!10^{\text{-}1}}$ \\ \hline
\multirow{2}{*}{12} & $-5.57\!\times\!10^{\text{-}2}$ & $ 2.06\!\times\!10^{\text{-}1}$ & $-2.65\!\times\!10^{\text{-}2}$ & $ 1.25\!\times\!10^{\text{-}1}$ \\
  & ${\it  4.83\!\times\!10^{\text{-}4}}$ & ${\it  1.95\!\times\!10^{\text{-}1}}$ & ${\it -3.03\!\times\!10^{\text{-}3}}$ & ${\it  1.16\!\times\!10^{\text{-}1}}$ \\ \hline
\multirow{2}{*}{13} & $-2.52\!\times\!10^{\text{-}2}$ & $ 2.13\!\times\!10^{\text{-}1}$ & $-1.11\!\times\!10^{\text{-}2}$ & $ 1.20\!\times\!10^{\text{-}1}$ \\
  & ${\it  5.21\!\times\!10^{\text{-}4}}$ & ${\it  1.88\!\times\!10^{\text{-}1}}$ & ${\it  2.68\!\times\!10^{\text{-}3}}$ & ${\it  1.12\!\times\!10^{\text{-}1}}$ \\ \hline
\multirow{2}{*}{14} & $ 3.69\!\times\!10^{\text{-}2}$ & $ 2.04\!\times\!10^{\text{-}1}$ & $-1.12\!\times\!10^{\text{-}2}$ & $ 1.17\!\times\!10^{\text{-}1}$ \\
  & ${\it -1.57\!\times\!10^{\text{-}3}}$ & ${\it  1.78\!\times\!10^{\text{-}1}}$ & ${\it  1.05\!\times\!10^{\text{-}3}}$ & ${\it  1.03\!\times\!10^{\text{-}1}}$ \\ \hline
\end{tabular}
\end{center}
\caption{This table lists the median and RMS variation in $\ra_{SDSS}-\ra$ and $\dec_{SDSS}-\dec$ for both CFHTLS-Wide and DEEP2, both before (regular text) and after ({\it italics}) the astrometric correction described in \S\ref{sec:astrom}. The RMS was calculated using a robust estimator of the standard deviation described in \S\ref{sec:cfhtcal}. There is significant improvement in both quantities for all pointings.}
\label{tab:astromcorr}
\end{table}



\begin{table}
\small
\begin{center}
\begin{tabular}{|c|cc|cc|cc|cc|cc|} \hline
\multirow{2}{*}{Pointing} & \multicolumn{2}{|c|}{$u$ band} & \multicolumn{2}{|c|}{$g$ band} & \multicolumn{2}{|c|}{$r$ band} & \multicolumn{2}{|c|}{$i$ band} & \multicolumn{2}{|c|}{$z$ band} \\
 & $a_0$ & $a_1$ & $a_0$ & $a_1$ & $a_0$ & $a_1$ & $a_0$ & $a_1$ & $a_0$ & $a_1$  \\ \hline
W3--1--2 &  0.0195 & -0.2309 &  0.0797 & -0.1691 &  0.0380 & -0.0106 &  0.0627 & -0.1538 &  0.0507 &  0.0427 \\
W3--1--3 & -0.0394 & -0.2043 &  0.0753 & -0.1770 &  0.0791 & -0.0210 &  0.0326 & -0.1452 &  0.0111 &  0.0985 \\
W3+0--1 & -0.0024 & -0.2010 &  0.1247 & -0.1877 &  0.1171 & -0.0379 &  0.0451 & -0.1422 &  0.0522 &  0.0630 \\
W3+0--2 & -0.0152 & -0.2334 &  0.0594 & -0.1673 &  0.0891 & -0.0221 &  0.0421 & -0.1480 &  0.0491 &  0.0619 \\
W3+0--3 & -0.0116 & -0.2320 &  0.0777 & -0.1687 &  0.0516 & -0.0107 &  0.0467 & -0.1611 &  0.0356 &  0.0832 \\
W3+1--1 &  0.0830 & -0.2149 &  0.0889 & -0.1820 &  0.0752 & -0.0193 &  0.0378 & -0.1394 &  0.0925 &  0.0677 \\
W3+1--2 &  0.0673 & -0.2198 &  0.0666 & -0.1756 &  0.0613 & -0.0276 &  0.0390 & -0.1494 &  0.0493 &  0.0730 \\
D3 & -0.0127 & -0.2247 &  0.0739 & -0.1695 &  0.0325 & -0.0194 &  0.0470 & -0.1495 &  0.0320 &  0.0893 \\ \hline \hline
$\langle a_1 \rangle$ &  & -0.2201 &  & -0.1746 &  & -0.0211 &  & -0.1486 &  &  0.0724 \\ \hline
\end{tabular}
\end{center}
\caption{Coefficients describing the linear relation between the magnitude difference in CFHTLS and SDSS (i.e. $m_c-m_s$) and the relevant SDSS color term for each CFHTLS-Wide pointing overlapping DEEP2, as well as for the CFHTLS-Deep pointing in that region. These were used in the CFHTLS photometric calibrations described in \S\ref{sec:cfhtcal}. The average value of $a_1$ is also listed for each band, which can be used to transform between the CFHTLS and SDSS photometric systems as described in \S\ref{sec:cfhtcal}.}
\label{tab:mcoeff}
\end{table}

\begin{table}
\small
\begin{center}
\begin{tabular}{|c|cc|cc|cc|cc|cc|} \hline
\multirow{2}{*}{Pointing} & \multicolumn{2}{|c|}{$u$ zero point offset} & \multicolumn{2}{|c|}{$g$ zero point offset} & \multicolumn{2}{|c|}{$r$ zero point offset} & \multicolumn{2}{|c|}{$i$ zero point offset} & \multicolumn{2}{|c|}{$z$ zero point offset} \\
 & median & $\sigma$ & median & $\sigma$ & median & $\sigma$ & median & $\sigma$ & median & $\sigma$  \\ \hline
\multirow{2}{*}{W3--1--2} & $ 1.31\!\times\!10^{\text{-}2}$ & $ 2.90\!\times\!10^{\text{-}1}$ & $ 7.97\!\times\!10^{\text{-}2}$ & $ 3.34\!\times\!10^{\text{-}2}$ & $ 3.83\!\times\!10^{\text{-}2}$ & $ 2.90\!\times\!10^{\text{-}2}$ & $ 6.31\!\times\!10^{\text{-}2}$ & $ 3.05\!\times\!10^{\text{-}2}$ & $ 4.99\!\times\!10^{\text{-}2}$ & $ 4.09\!\times\!10^{\text{-}2}$ \\
  & ${\it  1.33\!\times\!10^{\text{-}2}}$ & ${\it  2.90\!\times\!10^{\text{-}1}}$ & ${\it  1.98\!\times\!10^{\text{-}3}}$ & ${\it  3.25\!\times\!10^{\text{-}2}}$ & ${\it -4.89\!\times\!10^{\text{-}4}}$ & ${\it  2.82\!\times\!10^{\text{-}2}}$ & ${\it  2.21\!\times\!10^{\text{-}4}}$ & ${\it  3.01\!\times\!10^{\text{-}2}}$ & ${\it -1.51\!\times\!10^{\text{-}6}}$ & ${\it  4.05\!\times\!10^{\text{-}2}}$ \\ \hline
\multirow{2}{*}{W3--1--3} & $-5.55\!\times\!10^{\text{-}2}$ & $ 2.73\!\times\!10^{\text{-}1}$ & $ 7.47\!\times\!10^{\text{-}2}$ & $ 3.20\!\times\!10^{\text{-}2}$ & $ 8.01\!\times\!10^{\text{-}2}$ & $ 3.24\!\times\!10^{\text{-}2}$ & $ 3.30\!\times\!10^{\text{-}2}$ & $ 2.77\!\times\!10^{\text{-}2}$ & $ 1.26\!\times\!10^{\text{-}2}$ & $ 4.90\!\times\!10^{\text{-}2}$ \\
  & ${\it  8.42\!\times\!10^{\text{-}3}}$ & ${\it  2.73\!\times\!10^{\text{-}1}}$ & ${\it  9.25\!\times\!10^{\text{-}4}}$ & ${\it  3.13\!\times\!10^{\text{-}2}}$ & ${\it -7.68\!\times\!10^{\text{-}4}}$ & ${\it  3.09\!\times\!10^{\text{-}2}}$ & ${\it  5.73\!\times\!10^{\text{-}5}}$ & ${\it  2.65\!\times\!10^{\text{-}2}}$ & ${\it -1.37\!\times\!10^{\text{-}3}}$ & ${\it  4.81\!\times\!10^{\text{-}2}}$ \\ \hline
\multirow{2}{*}{W3+0--1} & $-1.08\!\times\!10^{\text{-}2}$ & $ 2.65\!\times\!10^{\text{-}1}$ & $ 1.26\!\times\!10^{\text{-}1}$ & $ 3.07\!\times\!10^{\text{-}2}$ & $ 1.16\!\times\!10^{\text{-}1}$ & $ 3.17\!\times\!10^{\text{-}2}$ & $ 4.54\!\times\!10^{\text{-}2}$ & $ 2.73\!\times\!10^{\text{-}2}$ & $ 5.33\!\times\!10^{\text{-}2}$ & $ 3.97\!\times\!10^{\text{-}2}$ \\
  & ${\it  6.03\!\times\!10^{\text{-}3}}$ & ${\it  2.65\!\times\!10^{\text{-}1}}$ & ${\it -1.21\!\times\!10^{\text{-}3}}$ & ${\it  3.06\!\times\!10^{\text{-}2}}$ & ${\it -1.04\!\times\!10^{\text{-}3}}$ & ${\it  3.12\!\times\!10^{\text{-}2}}$ & ${\it -2.60\!\times\!10^{\text{-}3}}$ & ${\it  2.57\!\times\!10^{\text{-}2}}$ & ${\it -1.64\!\times\!10^{\text{-}3}}$ & ${\it  3.94\!\times\!10^{\text{-}2}}$ \\ \hline
\multirow{2}{*}{W3+0--2} & $-7.93\!\times\!10^{\text{-}3}$ & $ 2.70\!\times\!10^{\text{-}1}$ & $ 6.00\!\times\!10^{\text{-}2}$ & $ 3.20\!\times\!10^{\text{-}2}$ & $ 8.92\!\times\!10^{\text{-}2}$ & $ 2.73\!\times\!10^{\text{-}2}$ & $ 4.17\!\times\!10^{\text{-}2}$ & $ 2.44\!\times\!10^{\text{-}2}$ & $ 4.97\!\times\!10^{\text{-}2}$ & $ 4.14\!\times\!10^{\text{-}2}$ \\
  & ${\it -2.67\!\times\!10^{\text{-}3}}$ & ${\it  2.70\!\times\!10^{\text{-}1}}$ & ${\it -5.26\!\times\!10^{\text{-}4}}$ & ${\it  3.09\!\times\!10^{\text{-}2}}$ & ${\it -6.18\!\times\!10^{\text{-}4}}$ & ${\it  2.65\!\times\!10^{\text{-}2}}$ & ${\it  6.71\!\times\!10^{\text{-}5}}$ & ${\it  2.33\!\times\!10^{\text{-}2}}$ & ${\it  9.10\!\times\!10^{\text{-}4}}$ & ${\it  4.05\!\times\!10^{\text{-}2}}$ \\ \hline
\multirow{2}{*}{W3+0--3} & $-2.10\!\times\!10^{\text{-}2}$ & $ 3.12\!\times\!10^{\text{-}1}$ & $ 7.70\!\times\!10^{\text{-}2}$ & $ 3.19\!\times\!10^{\text{-}2}$ & $ 5.13\!\times\!10^{\text{-}2}$ & $ 2.95\!\times\!10^{\text{-}2}$ & $ 4.74\!\times\!10^{\text{-}2}$ & $ 2.76\!\times\!10^{\text{-}2}$ & $ 3.58\!\times\!10^{\text{-}2}$ & $ 4.98\!\times\!10^{\text{-}2}$ \\
  & ${\it  1.90\!\times\!10^{\text{-}2}}$ & ${\it  3.12\!\times\!10^{\text{-}1}}$ & ${\it  1.02\!\times\!10^{\text{-}3}}$ & ${\it  3.19\!\times\!10^{\text{-}2}}$ & ${\it  1.64\!\times\!10^{\text{-}4}}$ & ${\it  2.92\!\times\!10^{\text{-}2}}$ & ${\it -7.70\!\times\!10^{\text{-}4}}$ & ${\it  2.76\!\times\!10^{\text{-}2}}$ & ${\it -7.94\!\times\!10^{\text{-}4}}$ & ${\it  4.87\!\times\!10^{\text{-}2}}$ \\ \hline
\multirow{2}{*}{W3+1--1} & $ 7.38\!\times\!10^{\text{-}2}$ & $ 2.49\!\times\!10^{\text{-}1}$ & $ 8.85\!\times\!10^{\text{-}2}$ & $ 3.09\!\times\!10^{\text{-}2}$ & $ 7.55\!\times\!10^{\text{-}2}$ & $ 3.18\!\times\!10^{\text{-}2}$ & $ 3.70\!\times\!10^{\text{-}2}$ & $ 2.87\!\times\!10^{\text{-}2}$ & $ 9.31\!\times\!10^{\text{-}2}$ & $ 4.80\!\times\!10^{\text{-}2}$ \\
  & ${\it  5.45\!\times\!10^{\text{-}3}}$ & ${\it  2.49\!\times\!10^{\text{-}1}}$ & ${\it  4.56\!\times\!10^{\text{-}4}}$ & ${\it  2.98\!\times\!10^{\text{-}2}}$ & ${\it  5.47\!\times\!10^{\text{-}4}}$ & ${\it  2.99\!\times\!10^{\text{-}2}}$ & ${\it -4.00\!\times\!10^{\text{-}5}}$ & ${\it  2.84\!\times\!10^{\text{-}2}}$ & ${\it -1.91\!\times\!10^{\text{-}3}}$ & ${\it  4.60\!\times\!10^{\text{-}2}}$ \\ \hline
\multirow{2}{*}{W3+1--2} & $ 5.27\!\times\!10^{\text{-}2}$ & $ 2.67\!\times\!10^{\text{-}1}$ & $ 6.54\!\times\!10^{\text{-}2}$ & $ 3.06\!\times\!10^{\text{-}2}$ & $ 6.14\!\times\!10^{\text{-}2}$ & $ 2.63\!\times\!10^{\text{-}2}$ & $ 3.82\!\times\!10^{\text{-}2}$ & $ 2.49\!\times\!10^{\text{-}2}$ & $ 5.08\!\times\!10^{\text{-}2}$ & $ 4.02\!\times\!10^{\text{-}2}$ \\
  & ${\it  8.09\!\times\!10^{\text{-}3}}$ & ${\it  2.67\!\times\!10^{\text{-}1}}$ & ${\it  3.69\!\times\!10^{\text{-}5}}$ & ${\it  2.99\!\times\!10^{\text{-}2}}$ & ${\it  7.76\!\times\!10^{\text{-}4}}$ & ${\it  2.57\!\times\!10^{\text{-}2}}$ & ${\it  2.09\!\times\!10^{\text{-}4}}$ & ${\it  2.45\!\times\!10^{\text{-}2}}$ & ${\it -4.76\!\times\!10^{\text{-}4}}$ & ${\it  3.92\!\times\!10^{\text{-}2}}$ \\ \hline
\multirow{2}{*}{D3} & $-1.77\!\times\!10^{\text{-}2}$ & $ 2.57\!\times\!10^{\text{-}1}$ & $ 7.43\!\times\!10^{\text{-}2}$ & $ 3.16\!\times\!10^{\text{-}2}$ & $ 3.13\!\times\!10^{\text{-}2}$ & $ 2.60\!\times\!10^{\text{-}2}$ & $ 4.60\!\times\!10^{\text{-}2}$ & $ 2.62\!\times\!10^{\text{-}2}$ & $ 3.23\!\times\!10^{\text{-}2}$ & $ 4.29\!\times\!10^{\text{-}2}$ \\
  & ${\it  4.66\!\times\!10^{\text{-}3}}$ & ${\it  2.57\!\times\!10^{\text{-}1}}$ & ${\it  7.02\!\times\!10^{\text{-}4}}$ & ${\it  3.16\!\times\!10^{\text{-}2}}$ & ${\it  8.28\!\times\!10^{\text{-}4}}$ & ${\it  2.60\!\times\!10^{\text{-}2}}$ & ${\it  1.39\!\times\!10^{\text{-}3}}$ & ${\it  2.62\!\times\!10^{\text{-}2}}$ & ${\it  6.27\!\times\!10^{\text{-}4}}$ & ${\it  4.29\!\times\!10^{\text{-}2}}$ \\ \hline
\end{tabular}
\end{center}
\caption{This table lists the median and RMS of the zero point offset relative to SDSS before (regular text) and after ({\it italics}) the calibration of the CFHTLS-Wide and Deep photometry. The RMS was calculated using a robust estimator of the standard deviation, as described in \S\ref{sec:cfhtcal}.  There is significant improvement in the median offset, such that it is much more consistent from pointing to pointing.  The decrease in the RMS is small for all pointings, as photometric errors for each individual object dominate over the systematic variation, but in essentially every case there is a detectable improvement from allowing the zero point correction to vary across the pointing (save for the $u$ band in Wide and all bands in Deep, where the corrections did not improve the RMS and hence have not been applied).}
\label{tab:zeropt}
\end{table}

\begin{table}
\small
\begin{center}
\begin{tabular}{|c|ccc|cc|ccc|cc|ccc|cc|} \hline
\multirow{2}{*}{Pointing} & \multicolumn{5}{|c|}{$B$ band} & \multicolumn{5}{|c|}{$R$ band} & \multicolumn{5}{|c|}{$I$ band} \\
 & $c_0$ & $c_1$ & $c_2$ & $d_0$ & $d_1$& $c_0$ & $c_1$ & $c_2$ & $d_0$ & $d_1$& $c_0$ & $c_1$ & $c_2$ & $d_0$ & $d_1$ \\ \hline
11 &  0.0507 &  0.4491 &  0.0895 & -0.0475 &  0.0149 &  0.0248 & -0.2358 &  0.0168 & -0.0185 &  0.0058 & -0.0069 & -0.3915 & -0.1944 & -0.0026 &  0.0004 \\
12 &  0.0987 &  0.2490 &  0.2290 & -0.0717 &  0.0216 &  0.0159 & -0.2211 &  0.0166 & -0.0359 &  0.0111 & -0.0068 & -0.3714 & -0.1796 &  0.0240 & -0.0085 \\
13 &  0.0366 &  0.5490 &  0.0073 & -0.0807 &  0.0221 &  0.0299 & -0.2557 &  0.0218 & -0.0467 &  0.0156 & -0.0038 & -0.4006 & -0.1404 &  0.0297 & -0.0104 \\
11 \& 12 &  0.0640 &  0.3843 &  0.1396 & -0.0639 &  0.0187 &  0.0239 & -0.2404 &  0.0246 & -0.0305 &  0.0094 & -0.0040 & -0.4083 & -0.1461 &  0.0144 & -0.0051 \\ \hline
\end{tabular}
\end{center}
\caption{Coefficients describing the transformation between CFHTLS-Wide $ugriz$ photometry and the DEEP2 $BRI$ system. The $c$ coefficients describe the quadratic fit for the color terms and the $d$ coefficients describe the linear fit of the object size-based correction, as described in \S\ref{sec:predphot}. The predicted $BRI$ photometry of pointing 14 was determined using the coefficients from the combination of objects in pointings 11 and 12.}
\label{tab:bricoeff}
\end{table}




\begin{table}
\footnotesize
\begin{center}
\begin{tabular}{cccccccccccccccc}
\hline\hline \\
OBJNO & RA\subscript{DEEP} & dec\subscript{DEEP} & RA\subscript{SDSS} & dec\subscript{SDSS} &
bestB & bestR & bestI & bestBerr & bestRerr & bestIerr & u & g & r & i & z \\
 & deg & deg & deg & deg & & & & & & & & & & & \\
\hline \\
11100002 & 214.33959 & 51.94544 & -99.0 & -99.0 & 24.8987 & 24.4452 & 24.2239 & 0.1611 & 0.1027 & 0.1935 & -99.0 & -99.0 & -99.0 & -99.0 & -99.0 \\
\\
11100003 & 214.35176 & 51.94724 & -99.0 & -99.0 & 24.9979 & 23.7502 & 23.5821 & 0.2009 & 0.0616 & 0.1219 & -99.0 & -99.0 & -99.0 & -99.0 & -99.0 \\
\\
11100004 & 214.35030 & 51.94721 & -99.0 & -99.0 & 25.5034 & 24.0699 & 23.6283 & 0.2600 & 0.0738 & 0.1034 & -99.0 & -99.0 & -99.0 & -99.0 & -99.0 \\
\\
11100000 & 214.36946 & 51.93833 & 214.36940 & 51.93835 & 25.2153 & 24.2380 & 23.6656 & 0.2787 & 0.1072 & 0.1404 & 24.7350 & 24.8243 & 24.8666 & 23.8106 & 25.0071 \\
\\
11100001 & 214.36935 & 51.93876 & 214.36935 & 51.93877 & 25.0578 & 24.5357 & 23.6953 & 0.1543 & 0.0903 & 0.0923 & 24.6641 & 24.4204 & 24.5461 & 23.3952 & 22.9106 \\
\\
11100008 & 214.27251 & 51.94721 & 214.27249 & 51.94729 & 24.5448 & 23.6794 & 22.9495 & 0.1433 & 0.0612 & 0.0722 & 25.1644 & 24.6349 & 24.3530 & 23.6916 & 23.2536 \\
\\
11100372 & 213.94919 & 52.07017 & 213.94912 & 52.07027 & 23.7849 & 23.5871 & 22.9274 & 0.0602 & 0.0573 & 0.0620 & 23.8947 & 23.2900 & 23.2600 & 22.7221 & 22.7198 \\
\\
11100402 & 213.64241 & 52.05337 & 213.64242 & 52.05340 & 24.3111 & 23.4869 & 23.0316 & 0.0737 & 0.0414 & 0.0547 & 25.0905 & 24.3498 & 23.9791 & 23.2209 & 23.1748 \\
\\
11100403 & 213.64160 & 52.05348 & 213.64157 & 52.05349 & 24.1674 & 23.4253 & 22.6158 & 0.0606 & 0.0368 & 0.0350 & 24.6405 & 24.0919 & 23.7302 & 22.9127 & 22.4641 \\
\\
\hline 
\\
\end{tabular}

\begin{tabular}{cccccccccccc}
\hline\hline \\
uerr & gerr& rerr & ierr & zerr & pgal & rg & zhelio & z\_err & zquality & SFD\_EBV & Source \\
 & & & & & & & & & & & \\
\hline \\
-99.0 & -99.0 & -99.0 & -99.0 & -99.0 & 3.00 & 2.59472 & -99.0 & -99.0 & -99 & 0.00854562 & DEEP \\
\\
-99.0 & -99.0 & -99.0 & -99.0 & -99.0 & 3.00 & 2.95285 & -99.0 & -99.0 & -99 & 0.00851513 & DEEP \\
\\
-99.0 & -99.0 & -99.0 & -99.0 & -99.0 & 3.00 & 2.39937 & -99.0 & -99.0 & -99 & 0.00851823 & DEEP \\
\\
0.1844 & 0.1591 & 0.2015 & 0.2003 & 0.6875 & 3.00 & 2.98541 & -99.0 & -99.0 & -99 & 0.00852105 & DEEP-CFHTLSW \\
\\
0.2191 & 0.1392 & 0.1903 & 0.1733 & 0.1265 & 1.00 & 1.91101 & -99.0 & -99.0 & -99 & 0.00851886 & DEEP-CFHTLSW \\
\\
0.1538 & 0.0756 & 0.0710 & 0.1015 & 0.0772 & 3.00 & 3.13192 & -99.0 & -99.0 & -99 & 0.00868888 & DEEP-CFHTLSW \\
\\
0.1357 & 0.0634 & 0.0749 & 0.1201 & 0.1356 & 3.00 & 2.95285 & 1.18247 & 6.40144e-05 & 4 & 0.0134441 & DEEP-CFHTLSW \\
\\
0.4416 & 0.0996 & 0.1601 & 0.0926 & 0.2277 & 3.00 & 2.38310 & 0.775875 & 4.99088e-05 & 4 & 0.0150038 & DEEP-CFHTLSW \\
\\
0.2792 & 0.0751 & 0.1218 & 0.0667 & 0.1132 & 3.00 & 2.23659 & 0.852072 & 4.44253e-05 & 4 & 0.0150446 & DEEP-CFHTLSW \\
\\ \hline
\end{tabular}
\end{center}
\caption{Examples of the catalog data for nine different objects in pointing 11: three objects with no matches, three with matches and no redshifts, and three with matches and redshifts.  Full data tables are provided at http://deep.ps.uci.edu/DR4/photo.extended.html.}
\label{tab:deepdata}
\end{table}


\end{document}